\documentclass[10pt,twocolumn]{IEEEtran}
\usepackage[numbers,sort&compress]{natbib}
\usepackage{amsmath,amssymb,epsfig}
\usepackage{fancybox}
\usepackage{tikz}
\newtheorem{definition}{\bf Definition}
\newtheorem{theorem}{\bf Theorem}
\newtheorem{corollary}{\bf Corollary}
\newtheorem{lemma}{\bf Lemma}

\begin{document}

\bibliographystyle{ieeetr}

\setlength{\parindent}{1pc}

\title{Explicit Estimators for Loss Tomography}
\author{Weiping~Zhu  \thanks{Weiping Zhu is with University of New South Wales, Australia, email w.zhu@adfa.edu.au}}
\date{}
\maketitle

\begin{abstract} In this paper, we unveil such a fact that the lack of explicit estimators in loss tomography is due to the lack of understanding the correlations considered by an estimator. By controlling the number and types of correlations, we can have a number of explicit estimators. To achieve this, the correlations that can be considered by an estimator are classified into different degrees. Based on the classification and composite likelihood, a number of explicit estimators are proposed in this paper that perform better than those proposed previously. The statistical properties of the explicit estimators, e.g. unbiasedness, consistency and asymptotic variance, are presented. In addition, a simulation study is conducted to confirm the properties obtained from analysis. Further, a number of explicit estimators based on composite likelihood are proposed for handling data missing in loss tomography.
\end{abstract}


\section{Introduction}

  Loss tomography has been studied for more than 10 years and a large number of estimators have been proposed during this period \cite{CDHT99, CDMT99, CDMT99a, CN00, XGN06, BDPT02,ADV07, DHPT06, ZG05, GW03}.
  Despite the overwhelming enthusiasm, almost all of the estimators proposed so far end up with a high degree polynomial, in particular for the maximum likelihood estimators (MLE),  that requires a numeric method, such as the Newton-Raphson algorithm, to find the solution of the polynomial.  The criticism of using a numeric method is its computation complexity that can affect the scalability and applicability of the estimators in practice. To address this criticism, a considerable effort has been made to find explicit estimators that have a similar performance as the MLEs.  Despite the effort, there has been little progress in this area. Only a handful explicit estimators are available that either perform worse than the MLE, e.g. \cite{DHPT06}, or do not have enough statistical analysis to support themselves although perform better than the estimator proposed in \cite{DHPT06}, e.g. \cite{Zhu11a}. All of those suggest  the need of a statistical principle  that can guide us to find efficient explicit estimators. This paper is devoted to provide such a principle that indeed leads us to a number of efficient explicit estimators.

  In order to have an explicit estimator as efficient as a numeric one, we carry out a thorough investigation of the statistical logic used by an maximum likelihood estimator
   (MLE). The investigation is focused on the correlations considered by the estimator since we believe the lack of explicit estimators is directly related to the lack of understanding the correlations embedded in the model of an estimator. The result received from the investigation confirms the belief and points out  the cause of using the numeric method in estimation is directly related to the number and type of correlations considered by an estimator. Based on the discovery, composite
   likelihood \cite{Lindsay88} is selected to create a number of likelihood functions that focus on a type or a few types of correlations.  The likelihood functions  subsequently lead us to a number of explicit estimators that have a  performance comparable to the MLE in terms of accuracy but better than the MLE in computation complexity.
We report the findings here that contribute to loss tomography in the following four-fold:
\begin{itemize}
\item the statistical logic used by a MLE is unveiled that provides a direction for us to look for explicit estimators;
\item composite likelihood  is selected as the tool to simplify the likelihood functions used previously in loss tomography.
    \item a large number of explicit estimators are proposed that makes estimator selection possible, i.e. select an estimator for a data set; and
    \item the statistical properties of the explicit estimators are studied and presented, including unbiasedness, consistency and uniqueness. In addition, the asymptotic variance of an explicit estimator is obtained that is comparable to that of the MLE.
    \end{itemize}

The rest of the paper is organised as follows. In Section \ref{section2}, we introduce the basics of loss tomography that include the loss model, the notations, and the statistics used in this paper. In Section \ref{section3}, we use the concepts introduced in Section \ref{section2} to have an MLE for a network of the tree topology. We then provide an analysis of the MLE that decomposes the estimator into a number of components. The decomposition throws light on the direction of explicit estimators.  Following the direction, we use composite likelihood to derive a number of likelihood equations in Section \ref{section4}. A statistic analysis of the proposed estimators is presented in Section \ref{section5} to discuss the unbiasedness, consistency and asymptotic variance of the estimators.
The estimators are further evaluated in Section \ref{section6} by simulations that confirm the contribution listed above. A brief discussion of the
 other features of the proposed estimators, including a number of explicit estimators for a data set with missing data, is also presented in this section.
Section \ref{section7} is devoted to concluding remark and future work.

\section{Assumption, Notation and Sufficient Statistics} \label{section2}

\subsection{Assumption}
To make loss tomography possible, probing packets, called probes, are multicasted from a source or a number of sources located on one side of a network to the receivers located on the other side of the network, where the paths connecting the sources to the receivers cover the links of interest. The multicast used here aims to create correlated observations at the receivers for statistical inference. If a network that does not support multicast, unicast-based multicast can be used to achieve the same effect \cite{HBB00},
\cite{CN00}. Statistical inference then uses the observation as the sample obtained from an experiment and puts it into a model that is assumed to generate the sample, where the pass (or loss) rate of a link or a path is the parameter to be estimated.
If the probes sent from sources to receivers are
far apart and network traffic remains statistically stable during the experiment, the observations are considered to be independent identical distributed
($i.i.d.$.) In addition to the probes, the losses occurred on a link or between links are assumed to be $i.i.d$.

\subsection{Notation}\label{treenotation}

 To make the following discussion rigorous, we use a large number of symbols in the following discussion that may overwhelm some of the readers who are not familiar with loss tomography. To avoid this, the symbols are gradually introduced through the paper. For  the frequently used symbols, we introduce them in this and next sections, whereas the others will be brought up later when needed.

Let $T=(V, E)$
be the multicast tree used to dispatch probes from a source to a number of receivers, where  $V=\{v_0, v_1, ... v_m\}$ is a set of
nodes and $E=\{e_1,..., e_m\}$ is a set of directed links that connect the
nodes in $V$. By default $v_0$ is the root node of the multicast tree to which the source is attached.  The set of leaf nodes $R, R \subset V$ represents all receivers attached to $T$. If $f(i)$ is used to denote
the parent of node $i$, there is a correspondence between nodes and links, where $e_i$ is the link connecting $v_{f(i)}$ to
$v_i$. $e_1$ is called the root link since it connects $v_0$ to $v_1$.

\begin{figure}
\begin{center}
\begin{tikzpicture}[scale=0.25,every path/.style={>=latex},every node/.style={draw,circle,scale=0.8}]
  \node            (b) at (25,30)  { $v_0$ };
  \node            (d) at (25,24) { $v_1$ };
  \node            (f) at (20,18)  { $v_2$ };
  \node            (g) at (30.5,18) { $v_3$ };
  \node            (j) at (16,12)  { $v_4$ };
  \node            (k) at (23,12) { $v_5$ };
  \node            (l) at (29,12) { $v_6$ };
  \node            (m) at (35,12)  { $v_7$ };
  \node            (r) at (12,6)  { $v_8$ };
  \node            (s) at (18,6) { $v_9$ };
  \node            (t) at (21,6) { $v_{10}$ };
   \node            (u) at (24.5,6)  { $v_{11}$ };
  \node            (v) at (27.5,6)  { $v_{12}$ };
  \node            (w) at (30.5,6) { $v_{13}$ };
  \node            (x) at (33.5,6) { $v_{14}$ };
  \node            (y) at (39,6) { $v_{15}$ };

  \draw[->] (b) edge (d);
  \draw[->] (d) edge (f);
  \draw[->] (d) edge (g);
  \draw[->] (f) edge (j);
  \draw[->] (f) edge (k);
  \draw[->] (g) edge (l);
  \draw[->] (g) edge (m);
  \draw[->] (j) edge (r);
  \draw[->] (j) edge (s);
  \draw[->] (k) edge (t);
  \draw[->] (k) edge (u);
  \draw[->] (l) edge (v);
  \draw[->] (l) edge (w);
  \draw[->] (m) edge (x);
  \draw[->] (m) edge (y);
\end{tikzpicture}
%\centerline{\psfig{figure=graph.eps,width=8.0cm,height=8.0cm}}
\caption{A Multicast Tree} \label{tree example}
\end{center}
\end{figure}
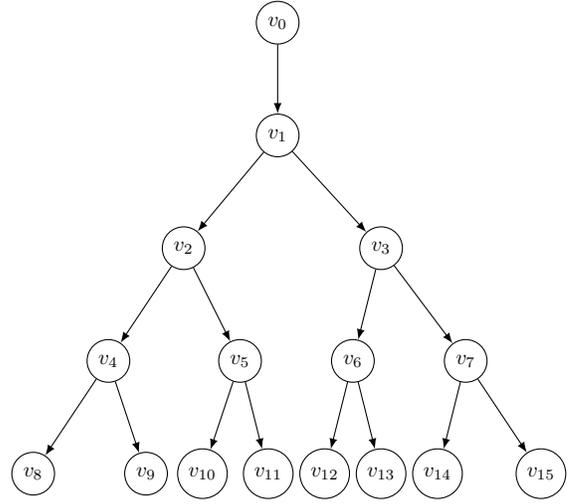

A multicast tree can be decomposed into a number of
 multicast subtrees, where $T(i)$ denotes the subtree that has $e_i$ as its root link and $R(i)$ denotes the receivers attached to $T(i)$.  In addition, we use $d_i$  to denote the descendants attached to node $i$ that is a nonempty set if $i \notin R$.  $|d_i|$ is used to denote the number of descendants in  $d_i$. In addition to $d_i$, $C_i$ is used to denote the multicast subtrees rooted at node $i$, called the children of node $i$. For example, Figure \ref{tree example} shows a complete binary multicast tree, where $R=\{v_8, v_9, \cdot\cdot, v_{15}\}$,  $R(2)=\{v_8, v_9, v_{10}, v_{11}\}$,  $d_{2}=\{4, 5\}$, $|d_{2}|=2$, and $C_{2}=\{T(4),T(5)\}$.

 If $n$ probes are sent from $v_0$ to $R$ in an experiment,
each of them gives rise of an independent realisation
of the passing (loss) process $X$. Let $X^{(i)}, i=1,...., n$ donate the $i-th$ process, where $X_k^i=1, k\in V$ if probe $i$
reaches $v_k$; otherwise $X_k^i=0$. The sample
$Y=(X_j^{(i)})^{i \in \{1,..,n\}}_{j \in R}$ comprises the observations of an experiment that can be divided into sections according to $R(k)$, where $Y_k$ denotes the part of $Y$ obtained by $R(k)$.
In addition to $Y_k$, we use $y_j^i$ to denote the observation of receiver $j,$ for probe $i$. $y_j^i=1$ if probe $i$ is observed by receiver $j$; otherwise, $y_j^i= 0$.

 Instead of using the pass rate of a link as the parameter to be estimated,  we use the pass rate of the path connecting $v_0$ to $v_k, k \in \{1,\cdot\cdot,m\}$ as the parameter, that is defined as the number of probes arrived at node $k$ divided by the number of probes sent from the source, i.e. $n$. Given the pass rates of all paths that connect $v_0$ to $v_k, k \in V\setminus v_0$, we are able to compute the pass rate of any link.  Let $\alpha_k$ be the pass rate of link $k$ and let $A_k$ be the pass rate of the path connecting $v_0$ to $v_k$. We have
 \[
 \alpha_k=\dfrac{A_k}{A_{f(k)}}.
 \]

\subsection{Sufficient Statistics}
\label{mlesection}

To estimate  $A_k, k \in \{1,\cdot\cdot,m\}$, we need to have a set of statistics for the likelihood functions of $A_k$s. The statistics are obtained from $Y$ and defined as follows:
\begin{equation}
 n_k(d_k)=\sum_{i=1}^n \bigvee_{\substack{l \in d_k \\ j \in R(l)}} y_j^i,  k \in \{1,\cdot\cdot,m\}.
 \label{nk1}
 \end{equation} \noindent $n_k(d_k)$ is a function of $d_k$ that returns the number of probes, confirmed from the receivers attached to $T(l), l \in d_k$, reaches node $k$. Thus, $n_k(d_k)$ is called the confirmed arrivals at node $k$ from $d_k$. Similarly, if  $x \subset d_k$, $n_k(x)$ returns the number of probes reaching the receivers attached to $T(j), j \in x$, and is called the confirmed arrivals at node $k$ from $x$.  In addition to $A_k$, we use $\beta_k$ to denote the pass rate of the subtree rooted at node $k$ and use $\gamma_k$ to denote the pass rate of a special multicast tree that is from $v_0$, via $v_k$, to $R(k)$. It is clear that $\gamma_k=A_k\cdot\beta_k, \forall k, k \in V \setminus (R\cup 0)$. If $k \in R$, we have $A_k=\gamma_k$. If $\hat\gamma_k$ denotes the empirical value of $\gamma_k$, we have $\hat\gamma_k=\dfrac{n_k(d_k)}{n}$.

 In addition to $C_k$, we introduce a concept called partial subtrees here that consist of a subset of $C_k$ and denoted by $C_k(x)$, where $x$ is the subset of the multicast trees rooted at node $k$. For instance, if $d_k=\{i,j,l\}$, $C_k(\{i,j\})=\{\{T(i), T(j)\}$ is the partial subtree consisting of $T(i)$ and $T(j)$. Given partial subtrees, we use $\beta_k(x), x \subset d_k$ to denote the pass rate of $C_k(x)$ and use $\gamma_k(x)=A_k\cdot \beta_k(x)$ to denote the pass rate from $v_0$, via $v_k$, to the receivers attached to $T(q), q \in x$. Similar to $\hat\gamma_k$, we have $\hat\gamma_k(x)=\dfrac{n_k(x)}{n}$, where
 \begin{equation}
 n_k(x)=\sum_{i=1}^n \bigvee_{\substack{l \in x \\ j \in R(l)}} y_j^i.
 \label{nkx}
 \end{equation}

Given $n_k(d_k)$,  we are interested in whether $n_k(d_k)$ is a sufficient statistic with respect to ({\it wrt.}) the passing process of $A_k$. Rather than using the well known factorisation theorem to confirm this, we directly use the mathematic definition of a sufficient statistic (See definition 7.18 in \cite{RM96}) to prove it. The definition {\it wrt.} the statistical model defined for the passing process is presented as a theorem here:

\begin{theorem}\label{complete minimal sufficient statistics}
Let $Y=\{X^{(1)},....,X^{(n)}\}$ be a random sample, governed by the
probability function $p_{A_k}(Y)$. The statistic $n_k(d_k)$ is minimal
sufficient for $A_k$.
\end{theorem}

\begin{IEEEproof}
According to the definition of sufficiency, we need to prove
\begin{equation}
p_{A_k}(Y|n_k(d_k)=t)=\dfrac{p_{A_k}(Y)}{p_{A_k}(n_k(d_k)=t,Y)}
\label{suff-condition}
\end{equation} is independent of $A_k$, which can be achieved in two steps.
\begin{enumerate}
\item According to the assumptions made previously, the passing process of a path is an {\it i.i.d.} process following the Bernoulli distribution. The probability function for observing $Y$ is as follows:
\begin{equation}
p_{A_k}(Y)=(A_k\beta_k)^{n_k(d_k)}(1-A_k\beta_k)^{n-n_k(d_k)}.
\label{likelihood0}
\end{equation}
\item Considering $Y$ the sample space that has $n_k(d_k)=t$, we have a binomial distribution
\[
p_{A_k}(n_k(d_k)=t, Y)=\binom{n}{t}(A_k\beta_k)^{t}(1-A_k\beta_k)^{n-t}.
\]
\end{enumerate}
Then, we have
\begin{eqnarray}
p_{A_k}(Y|n_k(d_k)=t)&=&\dfrac{(A_k\beta_k)^{t}(1-A_k\beta_k)^{n-t}}{\binom{n}{t}(A_k\beta_k)^{t}(1-A_k\beta_k)^{n-t}.
} \nonumber \\
=\dfrac{1}{\binom{n}{t}},
\end{eqnarray}
which is independent of $A_k$. Then, $n_k(d_k), k \in \{1,\cdot\cdot,m\}$ is a set of sufficient statistics.

Apart from the sufficiency,
 $n_k(d_k)$,  as defined in (\ref{nk1}), is a count of the probes reaching $R(k)$, where each probe is counted once and once only regardless of how many receivers observe it. Therefore,  $n_k(d_k)$ is a minimal sufficient statistic.
\end{IEEEproof}
\section{Problem Formulation and Analysis} \label{section3}

\subsection{Maximum Likelihood Estimator} \label{2.a}

Given (\ref{likelihood0}), we have a likelihood function as follows for the observation obtained by $R$
 \begin{equation}
 L(A_k)=A_k^{n_k(d_k)}(1-A_k\beta_k)^{n-n_k(d_k)}.
 \label{likelihood}
 \end{equation}
  A $\beta_k$ is dropped from (\ref{likelihood0}) since it simplifies the following discussion and does not affect both the sufficiency established for $n_k(d_k)$ and the following parametric estimation. Using the maximum likelihood principle, we derive a likelihood
equation by differentiating $\log L(A_k)$ {\it wrt.} $A_k$
and letting the derivatives be 0. We then have

\begin{eqnarray}
\dfrac{n_k(d_k)}{A_k}-\dfrac{(n-n_k(d_k))\beta_k}{1-A_k\beta_k}=0.
\label{likelihood equation}
\end{eqnarray}

\noindent  To represent $\beta_k$ in $A_k$, we use
\begin{equation}
1-\beta_k=\prod_{j \in d_k} (1-\dfrac{\gamma_j}{A_k}).
\label{beta-k}
\end{equation}
Solving $\beta_k$ from (\ref{beta-k}) and using it in (\ref{likelihood equation}), we have a likelihood equation as
\begin{equation}
1-\dfrac{n_k(d_k)}{n \cdot A_k}=\prod_{j \in d_k} (1-\dfrac{\gamma_j}{A_k}).
\label{realmle}
\end{equation}
Using $\gamma_k$ to replace $\dfrac{n_k(d_k)}{n}$ since the latter is the empirical value of the former, we have a likelihood as follows:
\begin{equation}
1-\dfrac{\gamma_k}{A_k}=\prod_{j \in d_k} (1-\dfrac{\gamma_j}{A_k})
\label{minc}
\end{equation}
that is identical to the MLE presented in \cite{CDHT99}. (\ref{realmle})  will be used in the following discussion because it explicitly connects observations to correlations.

\subsection{Problem and Direction}

 It is clear that (\ref{realmle}) is a polynomial having a degree equal to $|d_k|-1$. If $|d_k|$ is equal to or more than 6, there is no an explicit solution for $A_k$  according to  Abel-Ruffini theorem and Galois theory. Thus, a numeric method, such as Newton-Raphson method, is needed to approximate a solution.  For a univariate polynomial as (\ref{realmle}), there are a number of methods available that have been thoroughly studied. Despite this, a numeric method is inarguably slower than an explicit solution. Therefore, finding an explicit estimator that has its accuracy comparable to (\ref{realmle}) becomes the key issue of loss tomography and has attracted a consistent attention in the last 10 years.

It appears that the degree of the likelihood equation is determined by the number of descendants attached to a node. However, from the viewpoint of statistical inference, it is the result of the correlations considered by the likelihood equation. If all of the correlations among the descendants, from simple to complex, are considered, called full likelihood in literature, we have an equation as (\ref{realmle}). If we are able to control the number of correlations to be considered, we can have a number of likelihood equations that are solvable analytically.  In order to achieve this, we need to find the likelihood used by (\ref{realmle})  and then consider whether there is an alternative to remodel the passing process with less correlations.

\subsection{Correlations in Observation}

 To control the number of correlations in a likelihood equation, we need to find the correlations involved in  the likelihood equation, which requires us to decompose the likelihood equation according to correlations. Hopefully, the decomposition can divide  the likelihood equation into two parts:  the decomposed observations and the decomposed correlations. The former is called  {\it observed values} and the latter is called  {\it predictors} later in the paper because the latter predicts the former in estimation.  For instance, $\gamma_i\cdot \gamma_j/A_k, i, j \in d_k$ is a predictor of the portion of the observations that are simultaneously observed by  $R(i)$ and $R(j)$, i.e. at least one from $R(i)$ and another  from $R(j)$. Such an observation is also called a shared observation.

To achieve the goal set above,
 a $\sigma$-algebra, $S_k$, is created over $d_k$, where $\Sigma_k=S_k \setminus \emptyset$ is the non-empty sets in $S_k$, each of $\Sigma_k$ is for a correlation related to an observed value and a predictor.  If  the number of elements in a set is defined as the degree of the correlation, $\Sigma_k$ can be divided into $|d_k|$ exclusive groups, one for a degree of correlation. We name the groups $S_k(i), i \in \{1,\cdot\cdot,|d_k|\}$, where $i$ represents the degree of the correlation relating to the sets of the group. For example, if $d_k=\{i,j,k,l\}$, $S_k(2)=\{(i,j),(i,k),(i,l),(j,k),(j,l),(k,l)\}$.
  Let $\#(x), x \in \Sigma_k$ be the number of elements in $x$. We have $\forall y, y \in S_k(i): \#(y)=i$.

Given $\Sigma_k$, we are able to decompose $n_k(d_k)$ into observed values, one for a member in $\Sigma_k$.
 $n_k(d_k)$, as the minimal sufficient statistic, contains all information in $Y_k$, no more and no less. That implies each probe observed by $R(k)$ is counted once and once only in $n_k(d_k)$ regardless of how many receivers in $R(k)$ observe the probe, which is equal to eliminating various redundancy from the sum of $n_j(d_j), j \in d_k$.
 To explicitly express $n_k(d_k)$ by $n_j(d_j), j \in d_k$s, we need to have a function of $x, x \in \Sigma_k$, $I_k(x)$ to compute the redundancy in $x$,  that is  the number of probes simultaneously observed by the member of $x$ .

  Let $u_j^i$ be the observation of $R(j)$ for probe $i$, which is equal to
\[
u_j^i=\bigvee_{k \in R(j)} y_k^i.
\]
Then, we have
\[I_k(x)=\sum_{i=1}^n \bigwedge_{j \in x} u_j^i, \mbox{\vspace{1cm} } x \in \Sigma_k.\]
  If $\#(x)=1$,
\[
I_k(\{j\})=n_j(d_j), j \in d_k,
\]
otherwise, it is the shared observation of $x$. Then, we have

\begin{equation}
n_k(d_k)=\sum_{i=1}^{|d_k|}(-1)^{i-1}\sum_{x \in S_k(i)} I_k(x). \label{n_k value}
\end{equation}
(\ref{n_k value}) states that $n_k(d_k)$ is equal to a series of alternating adding and subtracting operations that ensure each probe observed by $R(k)$ is counted once and once only in $n_k(d_k)$.

\subsection{Correlations in  MLE}

Given (\ref{n_k value}), we are able to decompose (\ref{realmle}) into  a number of pairs consisting of observed values and predictors. The following theorem presents the detail and points out the principle used by (\ref{realmle}) in estimation.

\begin{theorem} \label{minctheorem}

\begin{enumerate}
\item (\ref{realmle}) is a full likelihood estimator that considers all of the correlations in $\Sigma_k$, and
\item (\ref{realmle}) consists of  observed values and their predictors, one for a member of $\Sigma_k$. The estimate obtained from (\ref{realmle})  is a fit that minimises an alternating differences between observed values and predictors.
    \end{enumerate}
\end{theorem}
\begin{IEEEproof}
(\ref{realmle}) is a full likelihood estimator because of the use of (\ref{beta-k}) in the derivation that covers all of the correlations in $d_k$. In addition,  we can prove this by finding a 1-to-1 correspondence between the observed values in $S_k$ and their predictors. We achieve this in three steps.
\begin{enumerate}
\item  If  we use (\ref{n_k value}) to replace $n_k(d_k)$ from the left hand side (LHS) of (\ref{realmle}),  the LHS becomes:
\begin{multline}
1-\dfrac{n_k(d_k)}{n\cdot A_k}= \\1-\dfrac{1}{n\cdot A_k}\big
[\sum_{i=1}^{|d_k|}(-1)^{i-1}\sum_{x \in S_k(i)}I_k(x)].
\label{nkexpansion}
\end{multline}
\item If we expand the product term located on  the RHS of (\ref{realmle}), we have:
\begin{equation}
\prod_{j \in d_k}(1-\dfrac{\gamma_j}{A_k})=1-\sum_{i=1}^{|d_k|}(-1)^{i-1}\sum_{x \in S_k(i)}\dfrac{\prod_{j \in x} \gamma_j}{A_k^i}. \label{prodexpansion}
\end{equation}
\item Deducting 1 from both (\ref{nkexpansion}) and (\ref{prodexpansion})
and then multiplying the results by $A_k$, (\ref{realmle}) turns to
\begin{eqnarray}
\sum_{i=1}^{|d_k|}(-1)^{i}\sum_{x \in S_k(i)}\dfrac{I_k(x)}{n}
=\sum_{i=1}^{|d_k|}(-1)^{i}\sum_{x \in S_k(i)}\dfrac{\prod_{j \in x} \gamma_j}{A_k^{i-1}}. \label{statequal}
\end{eqnarray}
It is clear there is a 1-to-1 correspondence between the terms across
the equal sign, where the terms on the LHS are the observed values and the terms on the RHS are the predictors. If we rewrite
(\ref{statequal}) as
\begin{equation}
\sum_{i=1}^{|d_k|}(-1)^{i}\sum_{x \in S_k(i)}\Big(\dfrac{I_k(x)}{n} -\dfrac{\prod_{j \in x}
\gamma_j}{A_k^{i-1}}\Big)=0, \label{correspondence}
\end{equation}
\end{enumerate}
the correspondence becomes obvious that shows the estimate obtained by (\ref{realmle}) is such an $A_k$ that ensures (\ref{correspondence}).
\end{IEEEproof}

\subsection{Insight of Full Likelihood Estimator} \label{insight}

 As expected, the pair within the bracket of the inner summation of (\ref{correspondence}) are the observed value of $x, x \in \Sigma_k$, and its predictor, respectively. The former is the empirical value of the latter.  By pairing them in an equation, we can have an estimator of $A_k$. Before validating this, we call such an estimator  a sub-estimator,  since it only considers a part of the correlations observed in estimation. The advantage of the sub-estimators, compared with the MLE, is its simplicity in expression, whereas the weakness is the same as the estimator presented in \cite{DHPT06}, i.e. it does not use all available information in observation that may affect the efficiency of the estimator.  If adding all sub-estimators that have the the same degree of correlation together, we can overcome the weakness in some degree, in particular if $A_k >90\%$. For instance, if the $\gamma_j$ occurred on the RHS of (\ref{statequal}) is replaced by its empirical value, i.e. $\hat\gamma_j=\dfrac{n_j(d_j)}{n}$, (\ref{statequal}) becomes
\begin{eqnarray}
\sum_{i=2}^{|d_k|}(-1)^{i}\sum_{x \in S_k(i)}\dfrac{I_k(x)}{n}
=\sum_{i=2}^{|d_k|}(-1)^{i}\sum_{x \in S_k(i)}\dfrac{\prod_{j \in x} \hat\gamma_j}{A_k^{i-1}} \label{statequal1}
\end{eqnarray}
since $\dfrac{I_k(\{j\})}{n}=\hat\gamma_j, j \in d_k$. This result shows the estimate of $A_k$ relies on the observed correlations to find a fit and each of them measures the fitness in its own dimension.
Pairing an inner summation on the LHS of (\ref{statequal1}) to its counterpart on the RHS, we can have $|d_k|-1$ estimators that are analytically solvable. Further,  selecting  a number of sub-estimators that have their correlations varying from $S_k(i)$ to $S_k(j)$ ($j-i<5$), we can have a number of explicit estimators that may further improve the efficiency of estimation . On the other hand, we can have a number of estimators by pairing an observed value with its predictor.
To prove the validity of the speculation, we need a different likelihood called composite likelihood to model the passing process of a multicast tree for $Y$.

\section{Composite Likelihood Estimator} \label{section4}
Composite likelihood is different from the full likelihood in the number of correlations considered. Instead of having all of the correlations, composite likelihood allows us to select some of the correlations in the likelihood function modelling the passing process from $v_0$ to $R(k)$. For completeness,  the likelihood is briefly introduced in \ref{intro composite}. Then,  a set of explicit estimators are proposed in this section. For those who are interested in the details of composite likelihood, please refer to \cite{Lindsay88}.
\subsection{Composite Likelihood} \label{intro composite}

 Composite likelihood dates back at least to the
pseudo-likelihood proposed by Besag \cite{Besay74}
for the applications that have large correlated data sets and highly
structured statistical model. Because of the complexity of the applications,  it is hard to explicitly describe the correlation embedded in observations.
Even if the correlation is describable, the computation
complexity often makes inference infeasible in practice \cite{VV05}. To overcome the difficulty, composite likelihood is proposed to handle those applications that only consider a number of the correlations within a given data set.
The composite likelihood defined in \cite{V08} is as follows:
\begin{definition} \label{compdefinition}
\begin{equation}
L_c(\theta;y)=\prod_{i\in I}f(y\in C_i;\theta)^{w_i}
\label{composite likelihood}
\end{equation}
where $f(y\in C_i;\theta)=f(\{y_j \in Y: y_j \in C_i\};\theta)$, with $y=(y_1,...,y_n)$, is a parametric statistical model,  $I \subseteq N$, and
$\{w_i, i \in I\}$ is a set of suitable weights. The associate composite log-likelihood
is $l_c(\theta;y)=\log L_c(\theta,y)$.
\end{definition}

 Using the maximum likelihood principle on a composite
likelihood function as (\ref{composite likelihood}), we have $\triangledown\log L_c(\theta,y)$, the composite likelihood equation. Solving the
composite likelihood equation, the maximum composite likelihood
estimate is obtained, which under the usual regularity condition
is unbiased and the associated maximum composite likelihood estimator
is consistent and asymptotically normally distributed \cite{XR11}.
%with mean equaling to true mean and variance matrix equaling to \[H(\theta)^{-1}J(\theta)H(\theta) \] where $J(\theta)$ is the variance of the composite likelihood equation and $H(\theta)$ is the expectation of the derivative of the composite likelihood equation.
Because of the properties, composite likelihood has
drawn considerable attention in recent years for the applications having complicated correlations, including spatial statistics \cite{ZJ10}, multivariate
survival analysis generalised linear mixed models, and so on \cite{ZJ09}.
Unfortunately, there has been little
work using composite likelihood in network tomography  although it is one of
the applications that have complex correlations. As far as we know, only \cite{LY03}
proposes an estimator on the basis of pseudo-likelihood for delay tomography and SD traffic matrix tomography.
%An estimator ofdelay tomography carries out a deconvolution process that is a difficult task since there are many options to turn an observed path delay into link ones and the options are restricted by the other observed delays. To estimate the delay of a number of serially connected links (called subpath), one must consider all impacts created by the descendants of the subpath if a full likelihood is used. However, this is almost impossible. \cite{LY03} instead only consider exclusive paired correlations in observations. Because of this, the authors are able to write down a composite likelihood function although \cite{LY03} calls it pseudo-likelihood. With the help of the likelihood function, a pseudo-likelihood EM algorithm is created and used in estimation.

\subsection{Explicit Estimators} \label{Explicit subsection}
Given composite likelihood, we are able to confirm the speculation presented in \ref{insight}. Starting from  $S_k(2)$ and ending at $S(|d_k|)$,  a number of composite likelihood equations can be derived, one for a degree of correlation.  The degree varies from pairwise to $|d_k|$-wise.  In order to have a unique formula for all of $L_c(i; A_k; y), i \in \{2,\cdot\cdot,|d_k|\}$,  we let the single-wise likelihood function be 1. In addition, let $L_c(i; A_k; y)$ be the  composite likelihood function for $i-wise$ correlation. Then, the following lemma shows $L_c(i; A_k; y)$ can be expressed uniformly.

\begin{lemma} \label{recursive corollary}
There are a number of composite likelihood functions, one for a degree of correlation, varying from pairwise to $|d_k|$-wise. The likelihood functions, i.e.  $L_c(i; A_k; y)$s have a unique form as follows:
\begin{eqnarray}
L_c(i; A_k; y) &=&\dfrac{\prod_{x \in S_k(i)}A_k^{n_k(x)}(1-A_k\beta_k(x))^{n-n_k(x)}}{\prod_{y \in S_k(i-1)}A_k^{n_k(y)}(1-A_k\beta_k(y))^{n-n_k(y)}}. \nonumber \\
&& i \in \{2,\cdot\cdot,|d_k|\}
\label{recursive form}
\end{eqnarray}
 \end{lemma}
\begin{IEEEproof}
The nominator on the RHS of (\ref{recursive form}) is the likelihood function considering the correlations from pairwise to $i$-wise inclusively and the denominator is the likelihood functions from single-wise to $(i-1)$-wise. Then, the quotient of them is the likelihood dedicated to the $i$-wise correlation.
\end{IEEEproof}
 Let $A_k(i)$ be the estimator derived from {\it i-wise} likelihood. Then,  we have the following theorem.
\begin{theorem} \label{all explicit}
Each of the composite likelihood
equations obtained from (\ref{recursive form}) is an explicit estimator of $A_k$  in the following form:
\begin{equation}
A_k(i)=\Big (\dfrac{\sum_{\substack{ x \in S_k(i)}}
\prod_{j \in x} \gamma_j}{\sum_{x \in S_k(i)}\dfrac{I_k(x)}{n}}{\Big )} ^{\frac{1}{i-1}},  i \in
\{2,.., |d_k|\}. \label{approximateestimator}
\end{equation}
\end{theorem}
\begin{IEEEproof}
Firstly, we change (\ref{recursive form}) into a log-likelihood function. We then differentiate the log-likelihood function {\it wrt} $A_k$ and let the derivative  be 0. Eventually, we  have a likelihood equation as (\ref{approximateestimator}).
\end{IEEEproof}
In the rest of the paper, we use $A_k(i)$ to refer the $i-wise$ estimator and $\hat A_k(i)$ to refer the estimate obtained by $A_k(i)$. The $i$ occurred in the brackets is called the index of the estimator.

Theorem \ref{all explicit} and lemma \ref{recursive corollary} show that we can have an estimator of
$A_k$ by matching a number of predictors having the same degree of correlation with the corresponded observed values. If we move one step further along this direction, we are able to have more explicit estimators, one for a correlation, not a degree, in $\Sigma_k$. The a corollary follows.
\begin{corollary}  \label{local estimator}
\begin{equation}
Al_k(x)=\Big(\dfrac{\prod_{j\in x} \gamma_j}{\dfrac{I_k(x)}{n}}\Big),     x \in \Sigma_k, \#(x) \geq 2,
\label{local estimator1}
\end{equation}
is an estimator of $A_k^{\#(x)-1}$.
\end{corollary}
\begin{IEEEproof} Let $\alpha_j$ be the pass rate of link $j$.
Since $\prod_{j \in x}\gamma_j$ is equal to $A_k^{\#(x)}(\prod_{j \in x}\alpha_j )\beta_k(x)$ and $\dfrac{I_k(x)}{n}$ is an estimate of $A_k (\prod_{j \in x}\alpha_j)\beta_k(x)$, putting them into (\ref{local estimator1}) we have the corollary.
\end{IEEEproof}

Comparing (\ref{approximateestimator}) with (\ref{local estimator1}), one is able to identify that $\hat A_k(i)$ is  type of mean from $Al_k(x), \#(x)=i$,  rather than an arithmetic one,  that has its nominator equal to the
sum of the nominators of $Al_k(x), \#(x)=i$ and its denominator equal to the sum of the denominators of $Al_k(x), \#(x)=i$. Because the mean resembles to power mean, we call it power-like mean.

\section{Properties of Composite Likelihood Estimator} \label{section5}

  Given the large number of explicit estimators presented in the last section, we are interested in the statistical properties of them, include  unbiasedness, consistency, and uniqueness. As Section \ref{section4}, we use a number of theorems and lemmas to present the properties.

 \begin{theorem} \label{local maximum}
$\hat Al_k(x)$ is an unbiased estimate of $A_k^{\#(x)-1}$.
\end{theorem}

\begin{IEEEproof}
 Let $\hat n_k(d_k)$ be the number of probes reaching $v_k$, let $z_j, j \in d_k$ be the pass rate of $T(j)$ and  $\overline{z_j}=\frac{n_j(d_j)}{\hat n_k(d_k)}$ be the sample mean of $z_j$. Similarly,  let $\overline{A_k}=\frac{\hat n_k(d_k)}{n}$. We then have
\begin{eqnarray}
E(Al_k(x))&=&E\Big(\dfrac{\prod_{j\in x} \hat\gamma_j}{\dfrac{I_k(x)}{n}}\Big) \nonumber \\
&=& E\Big(\dfrac{\prod_{j\in x} \dfrac{n_j(d_j)}{n}}{\dfrac{\sum_{i=1}^n \bigwedge_{j \in x} y^i_j}{n}} \Big )\nonumber \\
&=& E\Big(\dfrac{ (\dfrac{\hat n_k(d_k)}{n})^{\#(x)}\prod_{j\in x} \dfrac{n_j(d_j)}{\hat n_k(d_k)}}{\dfrac{\hat n_k(d_k)}{n} \dfrac{\sum_{i=1}^{\hat n_k(d_k)}\prod_{j \in x} z_j}{\hat n_k(d_k)}} \Big ) \nonumber \\
&=& E\Big((\dfrac{\hat n_k(d_k)}{n})^{\#(x)-1}\Big)E\Big(\dfrac{\prod_{j\in x} \overline{z_j}}{\prod_{j\in x} z_j}\Big) \nonumber \\
&=&E\Big ( (\dfrac{\hat n_k(d_k)}{n})^{\#(x)-1}\Big) E\Big ( \prod_{j \in x} \dfrac{\overline{z_j}}{z_j} \Big ) \nonumber \\
&=&E\Big (\overline{A_k}^{\#(x)-1}\Big)\prod_{j \in x}(E(\dfrac{\overline{z_j}}{z_j})) \nonumber \\
&=&E\Big (\overline{A_k}^{\#(x)-1}\Big)\prod_{j \in x}(E(\dfrac{\dfrac{\sum_{1}^{\hat n_k(d_k)} z_j}{\hat n_k(d_k)}}{z_j})) \nonumber \\
&=& A_k^{\#(x)-1}
\label{weighted mean}
\end{eqnarray}

\end{IEEEproof}
Theorem \ref{local maximum} reveals such a fact that although using a part of the observations obtained from an experiment, we can have an unbiased estimator of $A_k^{\#(x)-1}$.  To simplify the following discussion, we call $Al_k(x)$  the local estimator based on $x$ and the estimate obtained by $Al_k(x)$ the local estimate of $x$, denoted by $\widehat Al_k(x)$. Given theorem \ref{local maximum} we have
  \begin{corollary} \label{local unbiased}
$(\widehat Al_k(x))^{\frac{1}{\#(x)-1}}, x \in \Sigma_k$ is an unbiased estimate of $A_k$.
  \end{corollary}
  \begin{IEEEproof}
Bring the root operation into the proof of theorem \ref{local maximum}, we have the corollary.
 \end{IEEEproof}
Since $A_k(|d_k|)=Al_k(|d_k|)$, we can conclude that $A_k(|d_k|)$ is an unbiased estimator of $A_k$. 
For the other $A_k(i), 2 \leq i < |d_k|$, using the same procedure as that used in theorem \ref{local maximum}, we can prove the unbiasedness of them and have the following theorem.
\begin{theorem}
$\hat A_k(i), i \in \{2, \cdot\cdot, |d_k|\}$ are unbiased estimates of $A_k$.
\end{theorem}
\begin{IEEEproof}
To prove this theorem,  we need to prove
\begin{equation}
E\Big (\Big ( \sum_{x \in S_k(i)} \dfrac{\prod_{j \in x} \overline{z_j}}{\sum_{x \in S_k(i)} \prod_{j \in x} z_j} \Big )^{\frac{1}{i-1}} \Big )=1.
\label{unbiasedkey}
\end{equation}

As the proof of theorem \ref{local maximum}, we have
\begin{eqnarray}
&&E\Big (\Big (  \dfrac{\sum_{x \in S_k(i)}\prod_{j \in x} \overline{z_j}}{\sum_{x \in S_k(i)} \prod_{j \in x} z_j} \Big )^{\frac{1}{i-1}} \Big ) \\ \nonumber
&=&E\Big (\Big (  \dfrac{\sum_{x \in S_k(i)}\prod_{j \in x}\frac{1}{\hat n_k(d_k)}\sum_{i=1}^{\hat n_k(d_k)} z_i}{\sum_{x \in S_k(i)} \prod_{j \in x} z_j} \Big )^{\frac{1}{i-1}} \Big )  \\ \nonumber
&=& E\Big (\Big (  \dfrac{\sum_{x \in S_k(i)}\prod_{j \in x} z_j}{\sum_{x \in S_k(i)} \prod_{j \in x} z_j}  \Big )^{\frac{1}{i-1}} \Big ) \\ \nonumber
&=& 1.
\label{unbiasedkey1}
\end{eqnarray}
Thus, the theorem follows.
\end{IEEEproof}

To prove $\hat A_k(i)$ is a consistent estimate, we need to prove the consistency of the local estimators first. The following lemma provides this.

\begin{lemma} \label{two proofs}
\begin{enumerate}
\item $\widehat Al_k(x)$ is a consistent estimate of $A_k^{\#(x)-1}$; and
\item $\widehat Al_k(x)^{\frac{1}{\#(x)-1}}$ is a consistent estimate of $A_k$.
\end{enumerate}
\end{lemma}
\begin{IEEEproof}
We have the following two points to prove the two clauses of the lemma, one for each.
\begin{enumerate}
\item Theorem \ref{local maximum} shows that $\widehat Al_k(x)$ is equivalent to the first moment of $A_k^{\#(x)-1}$. Then, according to the law of large number, $\widehat Al_k(x)\rightarrow A_k^{\#(x)-1}$. \label{first proof}
    \item From the above and the continuity of $Al_k(x)$ on the values of $\gamma_j, j \in x$ and $I_k(x)/n$ generated as $A_k$ ranges over its support set, the result follows.
        \end{enumerate}
\end{IEEEproof}
Then, we have
\begin{theorem} \label{approximate consistent}
   $\hat A_k(i)$ is a consistent estimate of $A_k$.
\end{theorem}
\begin{IEEEproof}
  As stated,  $\hat A_k(i)$ is a power-like mean that is in between the smallest local estimate and the biggest one, i.e.
\begin{equation}
\min_{x \in S_k(i)}  \widehat Al_k(x)^{\frac{1}{i-1}} \leq \hat A_k(i) \leq  \max_{x \in S_k(i)} \widehat Al_k(x)^{\frac{1}{i-1}}.
\label{average mean1}
\end{equation}
Since all of $Al_k(x), x \in S_k(i)$ are consistent estimators, $A_k(i)$ is a consistent estimator.
\end{IEEEproof}

For the uniqueness of $A_k(i)$, we have.
\begin{theorem}
If
\[
\sum_{\substack{ x \in S_k(i)}} \prod_{j \in x} \hat\gamma_j < \sum_{x \in S_k(i)}\dfrac{I_k(x)}{n},
\]
there is only one solution in $(0,1)$ for $\hat A_k(i), 2 \leq i \leq |d_k|$.
\end{theorem}
\begin{IEEEproof}
Since the support of $\hat A_k(i)$ is in [0,1), we can reach this conclusion from (\ref{approximateestimator}).
\end{IEEEproof}
Similarly, there is a condition for the uniqueness of $Al_k(x), x \in \Sigma_k$.

 (\ref{realmle}) is proved to be the minimum variance unbiased estimator of $A_k$ in \cite{Zhu11}. We are interested here in the difference between (\ref{realmle}) and
  (\ref{approximateestimator}) in terms of variance. However, due to the complexity of (\ref{realmle}) and (\ref{approximateestimator}) we are not able to have  exact expressions for the variances at this moment. Instead of using exact expressions, we choose to use the asymptotic variances to compare them and a similar procedure as that used in \cite{DHPT06} to accomplish this task, i.e. using the delta method.

Using the delta method to obtain the asymptotic variance of $A_k(i)$, we need to have the covariance matrix for the variables in $A_k(i)$. The following lemma addresses this issue that has two parts  1) providing the formulae to compute the covariances, and 2) using the formulae to compute the approximate value of each covariance. Compared with lemma 1 in \cite{DHPT06}, the lemma can be viewed as the generalisation of lemma 1 in \cite{DHPT06} since it, apart from considering the covariance between $\gamma_i$ and $\gamma_j, i, j \in d_k$, also considers the covariances between $\dfrac{I_k(x)}{n}$s, where $\#(x) \geq 1$.

To compute the approximate values, we use a part of the results presented in \cite{DHPT06}, where the loss rates of the links from $v_0$ to $v_k$ and the loss rates of the links within the variables are used to express the covariance. To be consistent with the presentation in  \cite{DHPT06}, some of the symbols used in the proofs of theorem 5 in \cite{CDHT99} and lemma 1 in \cite{DHPT06} are adopted here, which are briefly introduced below. For those who want to know the details of the symbols and how they are derived, please refer to \cite{CDHT99, DHPT06}.  Among the adopted symbols,  $\bar\alpha_k = 1- \alpha_k$ is for the loss rate of link $k$, and $\|\bar \alpha\|=\max_{k \in E}|\bar\alpha_k|$. In addition, let $t_x=\sum_{j \in x} \bar \alpha_j, x \in \Sigma_k$ and $s_k=\sum_{j \in a(k)} \bar\alpha_j$, where $a(k)$ denotes the ancestor links of link $k$. Let $\xi_x=\dfrac{I_k(x)}{n}, \#(x)\geq 1$. Then,  we have the lemma:

 \begin{lemma} \label{approximate value}
 \begin{enumerate}
 \item $Cov(\xi_x,\xi_y)=\xi_{x\vee y} -\xi_x\xi_y$
  if $x, y \in \Sigma_k$ and $y \nsubseteq x$ or $x \nsubseteq y$; otherwise $Cov(\xi_x,\xi_y)=\xi_x(1-\xi_y)$ if $y \subseteq x$ and $x, y \in \Sigma_k$. \label{theoryresult}
 \item $Cov(\xi_x,\xi_y)=s_k+\bar{\alpha}_y+O(\|\bar\alpha\|^2)$ if $y \subseteq x$ and $x, y \in \Sigma_k$,  $Cov(\xi_x,\xi_y)=s_k+t_{x\wedge y}+O(\|\bar\alpha\|^2)$ if $x, y \in \Sigma_k$ and $y \nsubseteq x$ or $x \nsubseteq y$. \label{approximateresult}
     \end{enumerate}
 \end{lemma}
 \begin{IEEEproof} We first prove the clause \ref{theoryresult}), and then use  clause \ref{theoryresult}) and lemma 1 in \cite{DHPT06} to prove clause \ref{approximateresult}).

 \ref{theoryresult})
  Since the passing process follows the Bernoulli distribution, we have
  \begin{eqnarray}
  E(\xi_x \xi_y)
  &=& \Big[\prod_{j\in x \vee y} P(Y_j=1|X_{k}=1)\Big] P(X_{k}=1)\nonumber \\
  &=& \Big[ \prod_{j\in x \vee y}\dfrac{P(Y_j=1 \bigwedge X_{k}=1)}{P(X_{k}=1)}\Big]P(X_{k}=1)\nonumber \\
  &=&\dfrac{ \prod_{j\in x \vee y}\gamma_j}{A_{k}^{\#(x \vee y)-1}} \nonumber
  \end{eqnarray}
 if $x\nsupseteq y$ and $y\nsupseteq x$. Then,
  \begin{eqnarray}
  Cov(\xi_x\xi_y)&=& E(\xi_x\xi_y) - E(\xi_x)E(\xi_y) \nonumber \\
  &=&\dfrac{\prod_{j\in x \vee y}\gamma_j}{A_{k}^{\#(x\vee y)-1}}-\dfrac{\prod_{j\in x} \gamma_j}{A_k^{\#(x)-1}}\cdot \dfrac{\prod_{i\in y} \gamma_i}{A_k^{\#(y)-1}} \nonumber \\
  &=& \dfrac{\prod_{j \in x} \gamma_j \prod_{i \in y} \gamma_i}{A_k^{\#(x)+\#(y)-2}}\Big(\dfrac{1}{A_k^{\#(x\wedge y)+1}}-1 \Big ) \nonumber \\
  &=& \xi_x\xi_y\Big(\dfrac{1}{A_k^{\#(x\wedge y)+1}}-1 \Big ) \nonumber \\
  &=& \xi_{x\vee y} -\xi_x\xi_y
  \label{encluded}
  \end{eqnarray}
If $x, y \in \Sigma_k$ and $y \subset x$, we have
\begin{eqnarray}
E(\xi_x \xi_y)&=& E(\xi_x)=\xi_x \nonumber
\end{eqnarray}
and then
\begin{eqnarray}
  Cov(\xi_x,\xi_y)= \xi_x(1-\xi_y) \nonumber
  \end{eqnarray}
\ref{approximateresult}) According to Lemma 1 in \cite{DHPT06}, $A_k=1-s_k+O(\parallel \bar\alpha \parallel^2)$ and $\gamma_k=1-s_k+O(\parallel \bar\alpha \parallel^2)$, $\xi_x=A_k\prod_{j \in x}\dfrac{\gamma_j}{A_k}$, $\xi_x=1-s_k-t_x+O(\|\bar\alpha\|^2)$.
  Using them in (\ref{encluded}), we have
  \begin{eqnarray}
   Cov(\xi_x,\xi_y)&=&s_k+t_{x\wedge y} + O(\|\bar\alpha\|)^2 \nonumber
  \end{eqnarray}
  for $x, y \in \Sigma_k$ and $y \nsubseteq x$ or $x \nsubseteq y$.
  For $x, y \in \Sigma_k$ and $y \subset x \wedge \#(y)=1$. Then,
 \begin{align}
   C&ov(\xi_x,\xi_y)= \xi_x(1-\xi_y) \nonumber \\
   &= (1-s_k-t_x+O(\|\bar\alpha\|)^2)(1-(1-s_y+ O(\|\bar\alpha\|)^2)) \nonumber \\
   &= (1-s_k-t_x+O(\|\bar\alpha\|)^2)(s_k+\bar{\alpha}_y+O(\|\bar\alpha\|)^2) \nonumber \\
   &\doteq s_k+\bar{\alpha}_y+O(\|\bar\alpha\|)^2. \nonumber
  \end{align}
 \end{IEEEproof}
 Given lemma \ref{approximate value}, we are able to construct the covariance matrix of an estimate and compute the the asymptotic variance of the estimate.  The following theorem provides the asymptotic variance of $\hat A_k(i)$.

\begin{theorem} \label{asymtotic variance}
As $n \rightarrow \infty$, $n^{\frac{1}{2}}(\hat A_k(i)-A_k)$ has an asymptotically Gaussian distribution of mean 0 and variance $v_k^E(A_k(i))$, where $v_k^E(A_k(i))=s_k+ O(\parallel \bar\alpha \parallel^2)$.
\end{theorem}
\begin{IEEEproof}
Using the central limit theorem, we can prove $n^{\frac{1}{2}}(A_k(i)-A_k)$ is an asymptotically Gaussian distribution of mean 0 and variance $v_k^E$ as $n \rightarrow \infty$. We here focus on proving the variance of $A_k(i)$ is the same as that of the MLE to the first order in the pass rate of a link. Let
 \[
 D(i)=\{x |\#(x)=i, x \in \Sigma_k\}.
 \]
 that has
 \[
 |D(i)|=\binom {|d_k|}{i}
 \]
 elements.
By the delta method, $n^{\frac{1}{2}}(\hat A_k(i)-A_k)$ converges in distribution to a zero mean Gaussian random variable with variance
\[
v_k^E(A_k(i))=\triangledown A_k(i) \cdot C^E\cdot\triangledown A_k(i)^T\]
where $\triangledown A_k(i)$ is a derivative vector obtained as follows:
\begin{multline}
\triangledown A_k(i)=\Big ( \Big \{\dfrac{\partial A_k(i)}{\partial \xi_x}: x\in D(i)\Big \},  \Big \{ \dfrac{\partial A_k(i)}{\partial \gamma_j}: j \in D(1) \Big\}\Big ). \nonumber
\end{multline}
 $C^E$ is a square
covariance matrix in the form of
\begin{equation}
C^E=
\begin{pmatrix}
 \{Cov(\xi_x,\xi_y)\} & \{Cov(\xi_x, \gamma_j)\} \\
\{Cov(\xi_x, \gamma_j) \}& \{Cov(\gamma_j, \gamma_{j'})\}
\end{pmatrix}
\label{covariance}
\end{equation}
Each of the four $Cov(\cdot)$s in (\ref{covariance}) is a matrix, where $x,y \in D(i)$ and $j , j'\in D(1)$.
To calculate $\triangledown A_k(i)$ and derive the first order of $v_k^E$ in $\bar\alpha$, let $de(i)$ denote the denominator of $A_k(i)$  and $no(i)$ denote the nominator. In addition, let $no(i,j) \subset no(i), j \in d_k$ be the terms of $no(i)$ that have $\gamma_j$. Then,  we have
\begin{multline}
\triangledown A_k(i)=\dfrac{A_k(i)}{i-1}(\{\dfrac{-1}{de(i)}: j \in D(i) \}, \nonumber \\  \dfrac{no(i,j)}{\gamma_j\cdot no(i)}: j \in D(1)\})
\end{multline}
The number of terms in the denominator and the nominator of $A_k(i)$ is the same, i.e. $|de(i)|=|no(i)|$ and $|no(i,j)|=\binom{|d_k|-1}{i-1}$ and $var(\xi_x)=\xi_x(1-\xi_x)=s_k+t_x+O(\|\bar\alpha\|^2)$. Inserting the values and those obtained in Lemma \ref{approximate value} into $\triangledown A_k(i)$ and $C^E$, we have
\begin{multline}
\triangledown A_k(i)=\dfrac{1}{i-1}(\{\dfrac{-1}{|D(i)|}: j \in D(i) \}, \nonumber \\\{\dfrac{\binom{|d_k|-1}{i-1} }{|D(i)|}: j \in D(1) \} )+ O(\parallel \bar\alpha \parallel) \nonumber
\end{multline}
and

\begin{align}
C^E=&s_k+
\begin{pmatrix}
 \{t_{x\wedge y}: x, y \in D(i)\} & \{ \bar\alpha_j : j \in D(1) \} \\
 \{\bar\alpha_j, j \in D(1)\} & Diag\{ \bar\alpha_j : j \in D(1) \}
\end{pmatrix}
\nonumber \\ &+ O(\|\bar\alpha\|^2)
\label{covariance1}
\end{align}
  $\{t_{x\wedge y}: x, y \in D(i)\}$ is a symmetric matrix, where the diagonal entries are equal to $t_x, x\in D(i)$ since $t_{x\wedge x}=t_x$. The sum of each column in $\{t_{x\wedge y}: x, y \in D(i)\}$ is equal to
   \[ \sum_{j \in x}\binom{|d_k|-1}{i-1}\bar{\alpha}_j.
    \]
    The bottom left matrix has $|D(1)|$ columns, there are $\#(x)$ entries in a column that equal to $\bar{\alpha}_j, j \in x$, respectively. The top right matrix is the transpose of the bottom left one that has  $\binom{|d_k|-1}{i-1}$ entries in a column that all equal to $\bar{\alpha}_j$; the bottom right matrix is a diagonal matrix that has its diagonal entries equaling to $\bar{\alpha}_j, j \in D(1)$.  Let $M$ denote the the matrix in (\ref{covariance1}). We have $\triangledown A_k(i)\cdot M \cdot \triangledown A_k(i)^T = 0$ Then, the theorem follows.
\end{IEEEproof}
Compared with theorem 3 (i) in \cite{DHPT06}, theorem \ref{asymtotic variance} shows that the asymptotic variance of $A_k(i)$ is the same as that of the MLE, to the first order in the pass rate of a link.
\begin{corollary}
Under the same condition as that stated in theorem \ref{asymtotic variance}, the asymptotic variance of $Al_k(x)$ is equal to $s_k+O(\parallel \bar{\alpha}\parallel^2)$.
\end{corollary}
\begin{IEEEproof}
$Al_k(x)$ can be considered a special $A_k(i)$ that has a two-element $\bigtriangledown Al_k(x)$ and a four-element $C^E$. Using the same procedure as
that in the proof of theorem \ref{asymtotic variance}, we have the corollary.
\end{IEEEproof}
\subsection{Example}
We use an example to illustrate that a composite estimators has an asymptotic variance that is much closer to that obtained by the MLE than the explicit estimator proposed in \cite{DHPT06}. The setting used here is identical to that presented in \cite{DHPT06}, where $v_k$ has three children, numbered 1, 2, 3, with a pass
rate of $\alpha$ and the pass rate from $v_0$ to $v_k$ is also equal to $\alpha$.
Three estimators are considered, which are $A_k(2)$,
$A_k(3)$ and  (\ref{minc}). Note that

\[
A_k(3)=\Big(\dfrac{\gamma_1\gamma_2\gamma_3}{\dfrac{I_k(\{1,2,3\})}{n}}\Big)^{1/2}
\]
is the same as the explicit estimator proposed in \cite{DHPT06}.
On the other hand,
\begin{equation}
A_k(2)=\dfrac{\gamma_1\gamma_2+\gamma_1\gamma_3+\gamma_2\gamma_3}{\dfrac{I_k(\{1,2\})}{n}+\dfrac{I_k(\{1,3\})}{n}+\dfrac{I_k(\{2,3\})}{n}}
\label{twoAk} \end{equation} where
\begin{equation}
v_k^E(A_k(2))=\nabla A_k(2) C^E \nabla A_k(2)
\label{delta}
\end{equation}
and
 \[
 D(2)=\{x |\#(x)=2, x \in \Sigma_k\}.
 \] We then have $C^E$ that is the $2*|d_k|$ dimensional square
covariance matrix in the form of (\ref{covariance}).

Since $\xi_{\{1,2\}}=\dfrac{I_k(\{1,2\})}{n}$,
$\xi_{\{2,3\}}=\dfrac{I_k(\{2,3\})}{n}$, and $\xi_{\{1,3\}}=\dfrac{I_k(\{1,3\})}{n}$, we have
$\xi_{\{i,j\}}=\alpha^3, i, j \in \{1,2,3\}, i\neq j$ and $\gamma_l=\alpha^2, l \in \{1,2,3\}$.
Then, the top-left of $C^E$ is
\begin{equation}
\begin{pmatrix}
 \alpha^3(1-\alpha^3)& \alpha^6(\dfrac{1}{\alpha^2}-1)&\alpha^6(\dfrac{1}{\alpha^2}-1)\\
\alpha^6(\dfrac{1}{\alpha^2}-1)&\alpha^3(1-\alpha^3)&
  \alpha^6(\dfrac{1}{\alpha^2}-1)&\\
\alpha^6(\dfrac{1}{\alpha^2}-1)&\alpha^6(\dfrac{1}{\alpha^2}-1)&\alpha^3(1-\alpha^3)\\
\end{pmatrix}
\nonumber
\end{equation}
the top-right is
\begin{equation}
\begin{pmatrix}
\alpha^3(1-\alpha^2)&\alpha^3(1-\alpha^2)&\alpha^6(\dfrac{1}{\alpha^2}-\dfrac{1}{\alpha})\\
\alpha^3(1-\alpha^2)&\alpha^6(\dfrac{1}{\alpha^2}-\dfrac{1}{\alpha})&\alpha^3(1-\alpha^2)\\
\alpha^6(\dfrac{1}{\alpha^2}-\dfrac{1}{\alpha})&\alpha^3(1-\alpha^2)&\alpha^3(1-\alpha^2),
\end{pmatrix}
\nonumber
\end{equation}
the bottom-left is
\begin{equation}
\begin{pmatrix}
\alpha^3(1-\alpha^2)&\alpha^3(1-\alpha^2)&\alpha^5(\dfrac{1}{\alpha}-1)\\
\alpha^3(1-\alpha^2)&\alpha^5(\dfrac{1}{\alpha}-1)&\alpha^3(1-\alpha^2)\\
\alpha^5(\dfrac{1}{\alpha}-1)&\alpha^3(1-\alpha^2)&\alpha^3(1-\alpha^2),
\end{pmatrix}
\nonumber
\end{equation}
and the bottom-right is
\begin{equation}
\begin{pmatrix}
\alpha^2(1-\alpha^2)&\alpha^4(\dfrac{1}{\alpha}-1)&\alpha^4(\dfrac{1}{\alpha}-1)\\
\alpha^4(\dfrac{1}{\alpha}-1)&\alpha^2(1-\alpha^2)&\alpha^4(\dfrac{1}{\alpha}-1)\\
\alpha^4(\dfrac{1}{\alpha}-1)&\alpha^4(\dfrac{1}{\alpha}-1)&\alpha^2(1-\alpha^2).
\end{pmatrix}
\nonumber
\end{equation}
In addition, we have
\begin{equation}
\nabla A_k(2)=(\dfrac{-1}{3\alpha^2}, \dfrac{-1}{3\alpha^2},
\dfrac{-1}{3\alpha^2}, \dfrac{2}{3\alpha},
\dfrac{2}{3\alpha},\dfrac{2}{3\alpha}). \nonumber
\end{equation}
Using (\ref{delta}), we have
\begin{equation}
v_k^E(A_k(2))= \bar\alpha- \dfrac{2}{3}\bar{\alpha}^2+O(\|\bar\alpha\|^3).\nonumber
\end{equation}
Comparing this with the results presented in \cite{DHPT06} for $A_k(3)$ and $\hat A_k$ (MLE), where
\[ v_k^E(A_k(3))= \bar\alpha-\dfrac{\bar\alpha^2}{4}+O(\|\bar\alpha\|^3)
\]
 and
 \[
 v_k^E(\hat A_k)=\bar\alpha-\bar\alpha^2+O(\|\bar\alpha\|^3),
 \]
Clearly, $A_k(2)$ performs better than $A_k(3)$ in terms of asymptotic variance. Using asymptotic relative efficiency as a measure, $A_k(2)$ is more efficient than $A_k(3)$. This indicates such a fact that an estimator having more terms in its nominator and denominator is more efficient than another with less terms.

\section{Estimator Evaluation} \label{section6}
As discussed, there are three types of estimators: local, explicit, and full likelihood, that vary from simple to complex.
We can consider the local ones evolve to the explicit estimators which evolve to the full likelihood one although the discovery in practice is other way around.  The statistic logic used in the assumed evolution
is to use more information available in $Y_k$ to estimate $A_k$ since an estimator using more information tends to have a smaller variance. Because of this, $\hat A_k(i)$ is expected to have a smaller variance than $\hat Al_k(x)$s $\{x|x \in \Sigma_k, \#(x)=i\}$ and $\hat A_k$ is smaller than that of $\hat A_k(i)$. Despite this, as $n\rightarrow \infty$, there is litter difference between the estimates obtained by them although some coverage quicker than the others.

\begin{table*}[th]
  \centering
  \scriptsize
  \begin{tabular}{|l|r|r|r|r|r|r|r|r|r|r|}  \hline
 Estimators &\multicolumn{2}{|c|}{Full Likelihood} & \multicolumn{2}{|c|}{Pair Likelihood} & \multicolumn{2}{|c|}{Triple Likelihood} &\multicolumn{2}{|c|}{Pair local} & \multicolumn{2}{|c|}{Triple local}	\\ \hline
samples & Mean & Var &	Mean & Var &	Mean & Var	& Mean & Var &  Mean &	Var	\\ \hline
300&	0.0088&	1.59E-05&	0.0088&	1.59E-05&	0.0088&	1.64E-05&	0.0087&	 1.59E-05&	0.0087&	1.61E-05 \\ \hline
600&	0.0089&	1.12E-05&	0.0089&	1.12E-05&	0.0089&	1.13E-05&	0.0089&	 1.10E-05&	0.0088&	1.12E-05 \\ \hline
900&	0.0092&	7.76E-06&	0.0092&	7.82E-06&	0.0091&	7.84E-06&	0.0092&	 7.90E-06&	0.0092&	8.15E-06 \\ \hline
1200&	0.0095&	6.13E-06&	0.0095&	6.13E-06&	0.0094&	6.17E-06&	0.0095&	 6.16E-06&	0.0095&	5.97E-06 \\ \hline
1500&	0.0096&	4.55E-06&	0.0096&	4.55E-06&	0.0096&	4.80E-06&	0.0096&	 4.78E-06&	0.0096&	4.33E-06 \\ \hline
1800&	0.0096&	1.82E-06&	0.0096&	1.81E-06&	0.0096&	1.92E-06&	0.0097&	 1.92E-06&	0.0096&	1.90E-06 \\ \hline
2100&	0.0097&	3.14E-06&	0.0097&	3.11E-06&	0.0097&	3.14E-06&	0.0097&	 3.02E-06&	0.0097&	3.08E-06 \\ \hline
2400&	0.0100&	1.32E-06&	0.0100&	1.32E-06&	0.0100&	1.36E-06&	0.0100&	 1.29E-06&	0.0099&	1.28E-06 \\ \hline
2700&	0.0100&	1.72E-06&	0.0100&	1.72E-06&	0.0100&	1.74E-06&	0.0100&	 1.81E-06&	0.0100&	1.83E-06 \\ \hline
3000&	0.0102&	2.96E-06&	0.0102&	2.97E-06&	0.0102&	3.01E-06&	0.0102&	 3.04E-06&	0.0102&	2.95E-06 \\ \hline
4800&	0.0103&	1.74E-06&	0.0103&	1.74E-06&	0.0103&	1.74E-06&	0.0103&	 1.75E-06&	0.0103&	1.81E-06 \\ \hline
9900&	0.0099&	8.18E-07&	0.0099&	8.23E-07&	0.0099&	8.20E-07&	0.0099&	 8.05E-07&	0.0099&	8.60E-07 \\ \hline
\end{tabular}
  \caption{Simulation Result of a 8-Descendant Tree with Loss Rate=$1\%$}
  \label{Tab2}
\end{table*}

\begin{table*}
\centering
\scriptsize
\begin{tabular}{|l|r|r|r|r|r|r|r|r|r|r|}  \hline
 Estimators &\multicolumn{2}{|c|}{Full Likelihood} & \multicolumn{2}{|c|}{Pair Likelihood} & \multicolumn{2}{|c|}{Triple Likelihood} &\multicolumn{2}{|c|}{Pair local} & \multicolumn{2}{|c|}{Triple local}	\\ \hline
samples & Mean & Var &	Mean & Var &	Mean & Var	& Mean & Var &  Mean &	Var	\\ \hline
300&	0.0088&	1.59E-05&   0.0089&	1.64E-05&	0.0089&	1.68E-05&	0.0091&	 2.36E-05&	0.0088&	1.95E-05 \\ \hline
600&	0.0089&	1.12E-05&	0.0089&	1.14E-05&	0.0089&	1.16E-05&	0.0088&	 1.46E-05&	0.0089&	1.26E-05 \\ \hline
900&	0.0091&	7.76E-06&	0.0091&	7.80E-06&	0.0091&	7.83E-06&	0.0092&	 9.74E-06&	0.0091&	8.67E-06 \\ \hline
1200&	0.0094&	6.13E-06&	0.0094&	6.16E-06&	0.0094&	6.18E-06&	0.0096&	 7.09E-06&	0.0095&	6.16E-06 \\ \hline
1500&	0.0096&	4.55E-06&	0.0096&	4.72E-06&	0.0096&	4.81E-06&	0.0097&	 4.36E-06&	0.0096&	4.45E-06 \\ \hline
1800&	0.0096&	1.82E-06&	0.0096&	1.90E-06&	0.0096&	1.95E-06&	0.0096&	 2.45E-06&	0.0096&	1.97E-06 \\ \hline
2100&	0.0097&	3.14E-06&	0.0097&	3.11E-06&	0.0097&	3.11E-06&	0.0098&	 3.39E-06&	0.0097&	3.04E-06 \\ \hline
2400&	0.0099&	1.32E-06&	0.0100&	1.34E-06&	0.0100&	1.35E-06&	0.0101&	 1.64E-06&	0.0100&	1.44E-06 \\ \hline
2700&	0.0100&	1.72E-06&	0.0100&	1.69E-06&	0.0100&	1.67E-06&	0.0101&	 2.11E-06&	0.0100&	1.90E-06 \\ \hline
3000&	0.0102&	2.96E-06&	0.0102&	2.93E-06&	0.0102&	2.91E-06&	0.0103&	 2.83E-06&	0.0102&	2.87E-06 \\ \hline
4800&	0.0103&	1.74E-06&	0.0104&	1.74E-06&	0.0104&	1.74E-06&	0.0104&	 2.06E-06&	0.0104&	2.01E-06 \\ \hline
9900&	0.0099&	8.18E-07&	0.0099&	8.30E-07&	0.0099&	8.36E-07&	0.0099&	 9.78E-07&	0.0099&	9.11E-07 \\ \hline
\end{tabular}
  \caption{Simulation Result of a 8-Descendant Tree, 6 of the 8 have Loss Rate=$1\%$ and the other 2 have Loss Rate=$5\%$}
  \label{Tab3}
\end{table*}

\begin{table*}
\centering
\scriptsize
\begin{tabular}{|l|r|r|r|r|r|r|r|r|r|r|}  \hline
 Estimators &\multicolumn{2}{|c|}{Full Likelihood} & \multicolumn{2}{|c|}{Pair Likelihood} & \multicolumn{2}{|c|}{Triple Likelihood} &\multicolumn{2}{|c|}{Pair local} & \multicolumn{2}{|c|}{Triple local}	\\ \hline
samples & Mean & Var &	Mean & Var &	Mean & Var	& Mean & Var &  Mean &	Var	\\ \hline
300&	0.0503&	2.15E-04&	0.0504&	2.15E-04&	0.0505&	2.14E-04&	0.0508&	2.18E-04&	0.0505&	2.16E-04\\ \hline
600&	0.0503&	8.23E-05&	0.0503&	8.21E-05&	0.0503&	8.19E-05&	0.0504&	8.24E-05&	0.0503&	8.27E-05\\ \hline
900&	0.0511&	5.85E-05&	0.0511&	5.81E-05&	0.0511&	5.79E-05&	0.0512&	5.79E-05&	0.0512&	5.88E-05\\ \hline
1200&	0.0506&	4.93E-05&	0.0506&	4.97E-05&	0.0507&	4.99E-05&	0.0507&	4.85E-05&	0.0507&	4.93E-05\\ \hline
1500&	0.0502&	2.24E-05&	0.0502&	2.24E-05&	0.0502&	2.23E-05&	0.0503&	2.33E-05&	0.0502&	2.32E-05\\ \hline
1800&	0.0500&	3.89E-05&	0.0500&	3.85E-05&	0.0500&	3.83E-05&	0.0501&	3.91E-05&	0.0500&	3.94E-05\\ \hline
2100&	0.0507&	1.16E-05&	0.0507&	1.19E-05&	0.0507&	1.20E-05&	0.0507&	1.09E-05&	0.0507&	1.13E-05\\ \hline
2400&	0.0510&	1.40E-05&	0.0510&	1.43E-05&	0.0510&	1.44E-05&	0.0510&	1.40E-05&	0.0510&	1.43E-05\\ \hline
2700&	0.0507&	1.31E-05&	0.0507&	1.34E-05&	0.0507&	1.35E-05&	0.0508&	1.35E-05&	0.0507&	1.34E-05\\ \hline
3000&	0.0508&	6.65E-06&	0.0508&	6.98E-06&	0.0508&	7.14E-06&	0.0508&	6.79E-06&	0.0508&	6.85E-06\\ \hline
4800&	0.0498&	1.09E-05&	0.0498&	1.10E-05&	0.0498&	1.10E-05&	0.0498&	1.11E-05&	0.0498&	1.11E-05\\ \hline
9900&	0.0496&	5.35E-06&	0.0496&	5.38E-06&	0.0497&	5.40E-06&	0.0496&	5.48E-06&	0.0496&	5.48E-06\\ \hline
\end{tabular}
  \caption{Simulation Result of a 8-Descendant Tree, the loss rate of the root link=$5\%$, 4 of the 8 have Loss Rate=$1\%$ and the other 4 have Loss Rate=$5\%$}
  \label{Tab4}
\end{table*}

For a particular data set,  the estimate obtained by an estimator in (\ref{approximateestimator}) or an estimator in (\ref{local estimator1}) may be better than that obtained by (\ref{minc}).  If we want to find the better ones or the best one, we can use the composite Kullback-Leigh divergence to identify them, where Akaike information criterion (AIC)  \cite{A73} is used to determine which estimator is better. However, computing AICs for each of the estimators certainly increases the cost of estimation. Therefore, this approach should not be used unless necessary.

\subsection{Simulation}

To compare the performance of the three types of estimators, three rounds of simulations are conducted in various setting. Five estimators: the full likelihood, pairwise likelihood, triple-wise likelihood, a pair local,  and a triple local, are compared against each other in the simulation and the results are presented in three tables, from Table \ref{Tab2} to Table \ref{Tab4}. The number of samples used in the simulations varies from 300 to 9900 in a step of 300. For each sample size, 20 experiments with different initial seeds are carried out and the means and variances of the estimates obtained by each of the five estimators are presented in the tables for comparison. Due to the space limitation, we only present a part of the results in the tables, where all of the means and variance for the samples varying from 300 to 3000 are included. For the samples from 3300 to 9900, only two of them, i.e. 4800 and 9900, are presented.

 Table \ref{Tab2} is the results obtained from a tree with 8 children, where the loss rate of a child is set to 1\%.  The result shows when the sample is small, the estimates obtained by all estimators are drifted away from the true value that indicates the data obtained is not stable. Once the sample size reaches 2000, the estimates approach to the true value because the data is stabilised around the true value. All of the estimators achieve the same outcome with the increase of samples. Generally, with the increase of samples, the variance reduces slowly although there are a number of exceptions. As expected,  there is little difference between the estimators in terms of the means and the variance of the estimates if the loss rate of a path or a child is around 1\%.

To investigate the impact of different loss rates on estimation, another round simulation is carried out on the same network topology.
 The difference between this round and the previous one is the loss rates of the children connected to the path of interest, where 6 of the 8 children have their loss rates equal to $1\%$ and the other two have their loss rates equal to $5\%$. The two children
 selected by the paired local estimator have their loss rates equal to $1\%$ and $5\%$, respectively. Two of the three children of the triple local estimator have their loss rates equal to $1\%$ and the other has its loss rate equal to $5\%$. The results are presented in Table \ref{Tab3}. Compared Table \ref{Tab3} with Table \ref{Tab2}, there is little difference among the full likelihood, the pairwise likelihood and the triple-wise likelihood estimators in terms of the means and variances obtained by them except for the two local estimators if the sample size is smaller than 1000. The two local estimators have a  slightly higher variances than that obtained in the first round. This is because more samples are needed to get a stable observation for a network having higher loss rates than that having lower loss rates.
Comparing the two local estimators, we notice that the triple one performs a slightly better than the paired one in terms of variance. This indicates that a local estimator is sensitive to the number of children selected for estimation and the more the better since $I_k(x)$ is more stable than $I_k(y), y \subset x $, in particular if $n$ is small.

To further investigate the impact of loss rates on estimation, we  conduct another round simulation, where the loss rate of the path of interest is increased from $1\%$ to $5\%$, and the loss rates of four children are set to $5\%$ and the other four to $1\%$.  The two local estimators consider the observations obtained from the children that have $1\%$ loss rate. The result is presented in Table \ref{Tab4}, it differs from the previous two tables in the variances that are a magnitude higher than that of the previous two regardless of the estimators. This indicates a large variance is expected for a long path that traverses a number of serially connected links since the loss rate of a path is proportional to the number of links in the path. To reduce the variance, more probes need to sent.

%consider whether the estimate obtained from (\ref{minc}) can make all of the terms in the first summation of (\ref{correspondence}) zero. If so, the estimate obtained by (\ref{pair equation}) is equal to that obtained by (\ref{minc}). If not, the {\it i.i.d.} assumption fails to hold. Then, the two estimators are almost equal since each of them only meet a part of the {\it i.i.d.} assumption, where (\ref{minc}) ensures (\ref{correspondence}) as a whole hold and (\ref{pair equation}) ensures the first term of (\ref{correspondence}). (\ref{pair equation}) has obvious advantage over others, it releases the {\it i.i.d.} restriction that makes its estimate focusing on paired correlation only, in particular if $n\ll \infty$.

\subsection{Other Estimators}

Apart from the estimators presented in (\ref{approximateestimator}) and  (\ref{local estimator1}), there are still a large number of composite likelihood estimators for loss tomography. For instance, we proposed a number of estimators in \cite{Zhu11a} that divide the  children of a node into a number of groups and consider a group as a virtual child of the node. Then, using the number of probes observed by the receivers attached to each group as the statistic for the group, we can reduce the number of children attached to a node to a small number of virtual children. If the number of virtual children is less than 6, we can have an explicit estimator. For instance,  if $d_k$ is divided into two groups called $k_1$ and $k_2$, respectively. We have $n_{k}(d_{k_1})$ and $n_{k}(d_{k_2})$ as the confirmed arrivals at node $k$ from the observations of $R(k_1)$ and $R(k_2)$. Then, we have a composite likelihood estimator  as
\begin{equation} \label{group estimator}
1-\dfrac{\gamma_k}{A_k}=(1-\dfrac{\gamma_{k}(k_1)}{A_k})(1-\dfrac{ \gamma_{k}(k_2)}{A_k}).
\end{equation}
Using empirical probabilities $\hat\gamma_k=\dfrac{n_k(d_k)}{n}$,
$\hat\gamma_{k}(k_1)=\dfrac{n_{k}(d_{k_1})}{n}$, and
$\hat\gamma_{k}(k_2)=\dfrac{n_{k}(d_{k_2})}{n}$ in (\ref{group estimator}), we have an estimate of $A_k$ as
\[
\breve{A_k}=\dfrac{\hat \gamma_{k}(k_1)\hat \gamma_{k}(k_2)}{\hat \gamma_{k}(k_1)+\hat \gamma_{k}(k_2)-\hat \gamma_{k}}.
\]

Expanding (\ref{group estimator}) as that in expending (\ref{realmle}) in the proof of theorem (\ref{minctheorem}), we are able to find the estimator in fact is a partial pairwise estimator that focuses on the pairwise correlations between $\{x| x \in k_1\}$ and $\{y| y \in k_2\}$. Thus, we can selectively pair data with predictors as that used in \cite{LY03} to have some estimators that are simpler than the explicit estimators presented in (\ref{approximateestimator}).

\subsection{With Missing Data}

Loss rate estimation with a part of data missing has been considered in \cite{DHTWF02}, where an EM procedure is used to approximate an estimate that corresponds to the full likelihood. Given composite likelihood, this problem can be handled differently without the need to  model the missing data process as the method proposed in \cite{YZC11}. We can either
\begin{itemize}
\item select an estimator that is not related to missing data, or
\item add weight parameters to the likelihood function proposed in (\ref{composite likelihood}), where the likelihood objects involving missing data have a weight corresponding to the amount of  missing for the models of missing at random (MAR) and missing completely at random (MCAR). Let $AM_k(i)$ be the explicit estimator that takes into account of missing data in the $i$-wise correlations and let $w_x$ be the weight of missing in regard to $I_k(x), x \in S_k(i)$ that is inversely proportional to the amount of missing from the probes sent by the source. Using the same procedure as that used in \ref{Explicit subsection}, we have
    \begin{equation}
AM_k(i)=\Big (\dfrac{\sum_{\substack{ x \in S_k(i)}}
w_x\prod_{j \in x}\hat \gamma_j}{\sum_{\substack{ x \in S_k(i)}}w_x\dfrac{I_k(x)}{n}}{\Big )} ^{\frac{1}{i-1}},  i \in
\{2,.., |d_k|\}.
\end{equation}
as the explicit estimators for the samples with missing data.
\end{itemize}

\section{Conclusion and Future Works}\label{section7}

This paper aims at finding inspirations that can lead us to explicit estimators for loss tomography. In order to achieve this goal, we decompose the frequently used MLE into a number of components according to the correlations embedded into the estimator. The decomposition shows the MLE considering all of the correlations among the decedents attached to the path of interest and relying on the correlations to find an estimate. In addition, there is substantial redundancy in both observations and predictors, which shows each of the components could be an estimator on its own right. To prove this, composite likelihood is introduced to write the likelihood functions for each of the components and the sufficient statistic identified for the MLE is divided into a number of sectors, one for each component. Differentiating the composite likelihood functions and setting the derivatives to 0, we have a number of explicit estimators as expected. To compare with the MLE, the statistical properties of the explicit estimators are investigated and presented in the paper.  Apart from  the statistical properties, a series of simulations are conducted and the results are reported that show  the estimators proposed in this paper have a performance comparable to that of the full likelihood estimator. The composite likelihood can also be used in other topologies to simplify estimation and we will investigate the details and report them in the future.

%\bibliography{../globcom06/congestion}

%\bibliography{./congestion}

\begin{thebibliography}{10}

\bibitem{CDHT99}
R.~C\'{a}ceres, N.~Duffield, J.~Horowitz, and D.~Towsley, ``Multicast-based
  inference of network-internal loss characteristics,'' {\em IEEE Trans. on
  Information Theory}, vol.~45, 1999.

\bibitem{CDMT99}
R.~C\'{a}ceres, N.~Duffield, S.~Moon, and D.~Towsley, ``{Inference of Internal
  Loss Rates in the MBone },'' in {\em IEEE/ISOC Global Internet'99}, 1999.

\bibitem{CDMT99a}
R.~C\'{a}ceres, N.~Duffield, S.~Moon, and D.~Towsley, ``Inferring link-level
  performance from end-to-end multicast measurements,'' tech. rep., University
  of Massachusetts, 1999.

\bibitem{CN00}
M.~Coates and R.~Nowak, ``Unicast network tomography using {EM} algorthms,''
  Tech. Rep. TR-0004, Rice University, September 2000.

\bibitem{XGN06}
B.~Xi, G.~Nichailidis, and V.~Nair, ``Estimating netwrok loss rates using
  active tomography,'' {\em JASA}, 2006.

\bibitem{BDPT02}
T.~Bu, N.~Duffield, F.~Presti, and D.~Towsley, ``Network tomography on {General
  Topologies},'' in {\em SIGCOMM 2002}, 2002.

\bibitem{ADV07}
V.~Arya, N.~Duffield, and D.~Veitch, ``Multicast inference of temporal loss
  charateristics,'' {\em Performance Evaluation}, vol.~9-12, 2007.

\bibitem{DHPT06}
N.~Duffield, J.~Horowitz, F.~L. Presti, and D.~Towsley, ``Explicit loss
  inference in multicast tomography,'' {\em IEEE trans. on Information Theory},
  vol.~52, no.~8, 2006.

\bibitem{ZG05}
W.~Zhu and Z.~Geng, ``Bottom up inference of loss rate,'' {\em Journal of
  Computer Communications}, vol.~28, no.~4, 2005.

\bibitem{GW03}
D.~Guo and X.~Wang, ``Bayesian inference of network loss and delay
  characteristics with applications to tcp performance predication,'' {\em IEEE
  trans. on Signal Processing}, vol.~51, no.~8, 2003.

\bibitem{Zhu11a}
W.~Zhu, ``An efficient loss rate estimator in multicast tomography and its
  validity,'' in {\em IEEE International Conference on Communicaation and
  Software}, 2011.

\bibitem{Lindsay88}
B.~C. Lindsay, ``Composite likelihood method,'' {\em Contemporary Mathematics},
  vol.~80, 1988.

\bibitem{HBB00}
K.~Harfoush, A.~Bestavros, and J.~Byers, ``Robust identification of shared
  losses using end-to-end unicast probes,'' in {\em Technical Report
  BUCS-2000-013}, Boston University, 2000.

\bibitem{RM96}
R.~Mittelhammer, {\em Mathematical Statistics for Economics and Business},
  vol.~78.
\newblock Springer, 1996.

\bibitem{Besay74}
J.~Besag, ``Spatial interaction and the statistical analysis of lattice system
  (with discussion),'' {\em Journal of Royal Statistical Society}, vol.~36,
  1974.

\bibitem{VV05}
C.~Varin and P.~Vidoni, ``A note on composite likelihood inference and model
  selection,'' {\em Biometrika}, vol.~92, no.~3, 2005.

\bibitem{V08}
C.~Varin, ``On composite marginal likelihoods,'' {\em ASTA Advances in
  Statistical Analysis}, vol.~92, no.~1, 2008.

\bibitem{XR11}
X.~Xu and R.~Reid, ``On the robustness of maximum composite likelihood
  estimate,'' {\em Journal of Statistical Planning and Inference}, 2011.

\bibitem{ZJ10}
Z.~Jin, {\em Aspects of Composite Likelihood Inference}.
\newblock PhD thesis, University of Toronto, 2010.

\bibitem{ZJ09}
Y.~Zhao and H.~Joe, ``Composite likelihood estimation in multivariate data
  analysis,'' {\em Canadian Journal of Statistics}, vol.~33, 3 2009.

\bibitem{LY03}
G.~Liang and B.~Yu, ``Maximum pseudo likelihood estimation in network
  tomography,'' {\em IEEE trans. on Signal Processing}, vol.~51, no.~8, 2003.

\bibitem{Zhu11}
W.~Zhu and K.~Deng, ``Loss tomography from tree topologies to general
  topologies,'' {\em arXiv:1105.0054}, 2011.

\bibitem{A73}
H.~Akaike, ``A new look at the statistical model identification,'' {\em IEEE
  Transactions on Automatic Control}, vol.~19, no.~6, 1974.

\bibitem{DHTWF02}
N.~Duffield, J.~Horowitz, D.~Towsley, W.~Wei, and T.~Friedman,
  ``Multicast-based loss inference with missing data,'' {\em IEEE Journal on
  Selected Area in Communication}, vol.~20, no.~4, 2002.

\bibitem{YZC11}
G.~Yi, L.~Zeng, and R.~Cook, ``A robust pairwise likelihood method for
  incomplete longitudinal binary data arising in cluster,'' {\em The Canadian
  Journal of Statistics}, vol.~39, no.~1, 2011.

\end{thebibliography}
%&=& E\Big(\dfrac{\prod_{j\in x} \dfrac{n_j(d_j)}{n}}{\dfrac{\sum_{i=1}^n \bigwedge_{j \in x} y_j^i}{n}}\Big ) \nonumber \\
%&=& E\Big(\dfrac{\prod_{j\in x}\Big(\dfrac{\dfrac{n_i(1)}{\hat n_k(d_k)}}{\dfrac{n}{\hat n_k(d_k)}}\Big)}{\dfrac{\sum_{i=1}^n f(z_x)^i \prod_{j \in x} z %_j^i}{n}} \Big ) \nonumber \\
%&=&  E\Big(\dfrac{\prod_{j \in x} \dfrac{n_j(d_j)}{\hat n_k(d_k)}}{\prod_{j \in x} \dfrac{n}{\hat n_k(d_k)}\cdot \dfrac{1}{n}\cdot n \cdot A_k \prod_{j \in %x}  \dfrac{n_j(d_j)}{\hat n_k(d_k)}} \Big ) \nonumber \\

\end{document}